\documentclass[twocolumn,twoside,slac]{revtex4}
\usepackage{graphicx}
\usepackage{fancyhdr}
\begin{document}

\title{IGUANA Architecture, Framework and Toolkit for Interactive Graphics}

%

\author{George Alverson, Giulio Eulisse, Shahzad Muzaffar, \\
Ianna Osborne, Lassi A. Tuura, Lucas Taylor}
\affiliation{Northeastern University, Boston, USA}

\begin{abstract} 

IGUANA is a generic interactive visualisation framework based on a C++
component model. It provides powerful user interface and visualisation
primitives in a way that is not tied to any particular physics experiment or
detector design. The article describes interactive visualisation tools built
using IGUANA for the CMS and  D0 experiments, as well as generic GEANT4 and
GEANT3 applications. It covers features of the graphical user interfaces, 3D and 2D graphics,
high-quality vector graphics output for print media, various textual, tabular and
hierarchical data views, and integration with the application through control
panels, a command line and different multi-threading models.
\end{abstract}

\maketitle

\thispagestyle{fancy}


\section{INTRODUCTION}

IGUANA - Interactive Graphics for User ANAlysis - is a modular C++
toolkit for interactive visualisation. The project mainly focuses on
interactive detector and event visualisation with emphasis on:
\begin{itemize}
\item High performance 2D and 3D graphics;
\item Graphical user interfaces;
\item User access to the experiment services: data access framework,
  application execution framework, etc.
\end{itemize}

The goal of the project is to provide easy-to-use coherent interactive
graphical application interface for the physicist, where the main
application - IGUANA studio - interfaces to other tools and components.
Interactive histogram, ntuple analysis is not considered a primary
goal. It is assumed that this functionality will be provided by other
tools and projects such as JAS~\cite{Jas}, Hippodraw~\cite{hippo-01},
ROOT~\cite{Root}, or OpenScientist~\cite{Opensci}, which may (or may
not) ultimately be integrated with IGUANA. 

IGUANA development is mainly driven by the needs of the CMS and D0
experiments, but most of IGUANA is independent of CMS or D0 and can be freely
used by others. In CMS, IGUANA releases are available directly via SCRAM~\cite{scram}
which keeps a list of the IGUANA project releases. Instructions
versioned by release on how to download and install IGUANA are
available for any selected version. 

IGUANA is currently supported on Linux, Solaris and Windows. Porting
to other unix varieties is reasonably straightforward. 

\section{ARCHITECTURE AND FRAMEWORK}
 
IGUANA architecture is designed to provide a generic tool that can
integrate with a specific experiment or even a single task within it
without having to specify a fixed list of abstract interfaces or
having to standardise on a common object description. IGUANA has a
small kernel, everything else is implemented in demand-loaded
extensions which negotiate the specific services they need. This
allows new services to be added quickly, to be propagated into
existing code gradually, and existing services to be altered to do new
things easily.

The full documentation is available on the project
website\footnote{http://iguana.cern.ch} automatically generated for
every release from the sources kept in CVS.

\subsection{Core Overview} \label{Plgin}

\begin{figure*}[t]
\centering
\includegraphics[width=155mm]{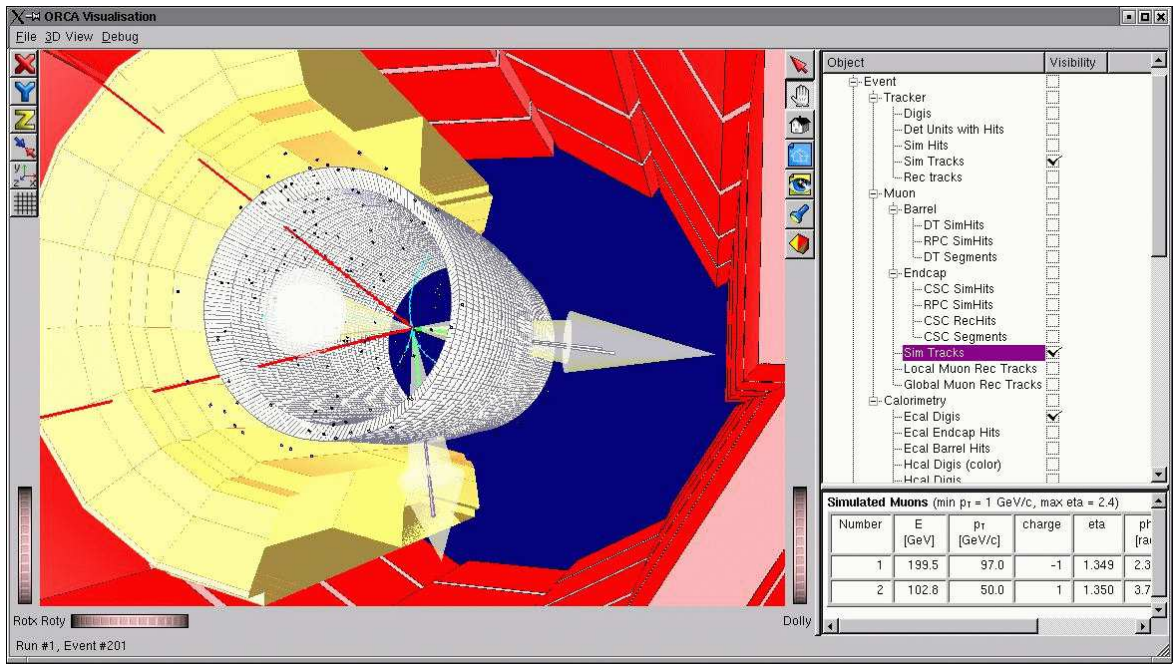}
\caption{ORCA visualisation based on IGUANA: IgSplitter sites
  (horisontal and vertical) host the 3D browser, the Twig browser, and
  the Text browser. Communication is established among all the browsers:
  correlated picking from the 3D browser window broadcasts
  selection. The Twig browser responds by highlighting corresponding Twig
  leaf and the text rep in the form of an html text is displayed in the
  Text browser.} \label{orca-f3}
\end{figure*}

The main units relating to the IGUANA architecture are:

\begin{itemize}
\item A thin portability and utilities layer; 
\item A small kernel that manages a number of plug-ins: 
\begin{itemize}
\item application personalities; 
\item session with extensions forming the shared application state; 
\item user interface components: sites, browsers; 
\item representation methods to map between the experiment objects and
  the various browsers; 
\end{itemize}
\item  External software imported into IGUANA for convenience of
  building and distribution, and external software which remains outside IGUANA.
\end{itemize}
                                  
The portability layer, is mostly implemented by the classlib package
which in future will be distributed by the LHC Computing Grid (LCG)
SEAL project~\cite{Seal}.

The kernel is implemented by IgPluginManager, a generic
plug-in manager package, which is extended by the architecture kernel
package IgObjectBrowser that provides the core functionality: session
objects, extensions, user interface components, representations, etc.
and base interfaces for the various plug-in types.

The plug-ins are in the Ig\_Modules subsystem which includes the event
display core, application management services, and thin adaptors to
export external software in forms understood by the architecture. For
more details see~\cite{arch-01}.

\subsection{Application Personalities}

When an IGUANA program runs, it first creates a session object into
which it then attaches an application personality: the main program
that determines all subsequent behaviour (IgSession,
IgDriver). Typically the personality immediately extends the session 
object with services pertinent to the purpose of the application
(IgExtension). For example, a graphical personality would create a
main window and add services that give access to the menus and
maintain GUI event loops. The personality then loads a number of
extensions into the session; a graphical personality would tell the
system to load all ``GUI extensions'' (IgExtension). 

The personality and the extensions form the application. The
personality exposes features by installing service interfaces into the
session, based on which other extensions can provide further services and
make features available to the user.

\subsection{Browsers and Sites}

An interactive application personality creates its user
interface based on one of the user interface components: browsers and
sites (IgBrowser, IgSite). A browser is a way to look at and interact
with application objects (see Figure~\ref{orca-f3}). It may also control other things,
such as a cut in a histogram display. A browser does not however have
to be graphical: it might just sit on the background and respond to other
browsers' queries such as ``What actions can be invoked on this
object?'' Sites host browsers, by providing for example a window for a
3D browser view. More generally sites expose objects such as GUI
widgets as hosts for other sites and browsers. Sites need not to be
graphical either: a pipe site might host a browser communicating to a
subprocess.

\subsection{Object Representations}

Application objects can be represented in many different browsers. A
browser typically has a model made of a number of object reps
(IgModel, IgRep). The use of model-specific reps, such as
3D shape objects, is encouraged rather then the application objects
themselves. This permits an object's representation to be separated
from the object itself. It is possible to correlate the object's reps
in different browsers and to create a rep in a right context in each
browser.

This implementation requires ``representable'' application objects to
inherit from a common base type (IgRepresentable) or make it an alias
for another class. The
only requirement is that the type must be polymorphic; IgRepresentable
simply defines a virtual destructor and no other methods. 

In CMS there is however no base type shared by all objects; for
visualisation the proxies are used that inherit IgRepresentable and
point to other CMS objects. The primary reason for this choice is the
framework's support for ``virtual objects'' that can be materialized
upon request. 

The object -- rep mappings are extensions loaded dynamically on
demand. IGUANA discovers and chooses the right mapping automatically;
the extension code simply does whatever is appropriate for that
combination. It is not necessary to have a global list of all the
pairings: code for new reps and views can be dropped in without any
changes to existing code.

\subsection{Communication}

Communication takes place through three channels:
session extensions, message services, and the object mapping methods.
The first has already been covered. The message services allow
browsers to share messages such as ``I selected this'': all observers
of a service can maintain together a coherent state and to answer
queries from each other while still knowing next to nothing about the
message sender (see Figure~\ref{orca-f3}). 

The final channel is the object--rep--model
operations, one of which was already mentioned: the creation of a
rep. Another operation commits a changed rep back to the object, yet
another lazily expands an object rep (for example to read data only
when it is requested). The operations can be extended, both to handle
new argument combinations and with new methods with arbitrary
parameter types (IgBrowserMethods).

\section{VISUALISATION FRAMEWORK FOR GEANT4}\label{geant4}

IGUANA provides an interactive visualisation sub-framework for
GEANT4~\cite{Geant4}. The framework implements 
DCUT, DTREE-like functionality (as in Geant3). It allows to
explore and visualise the volume tree, with all the usual IGUANA 3D
features: view, correlated picking, slices per object, etc. with
the Geant4 command line still accessible.
Navigation in the volume tree can be done either by logical or
physical volumes, or subsets. There are quick operations for common tasks.
Volume property window shows additional information about selected
object or volume.
Volume tree selectors provide various types of selection:

\begin{itemize}
\item by material: "show all silicon";
\item by sensitive: "show only sensitive detectors";
\item for a sub-tree, predefined viewpoints/settings;
\item forward + reverse: "show where this is used".
\end{itemize}

The framework is used in the CMS OSCAR simulation project where
an IGUANA wizard is used to guide through GEANT4/OSCAR settings.
The application is integrated with Martin Liendl's overlap detection:
overlaps are found and the result details are shown in a list.

Two generic examples showing how to use an arbitrary GEANT4 geometry with the
framework are distributed with IGUANA experiment-independent core: the
GEANT4 example number two and the ATLAS calorimeter example.

\section{APPLICATIONS}

\begin{figure*}[!hbtp]
\centering
\includegraphics[width=135mm]{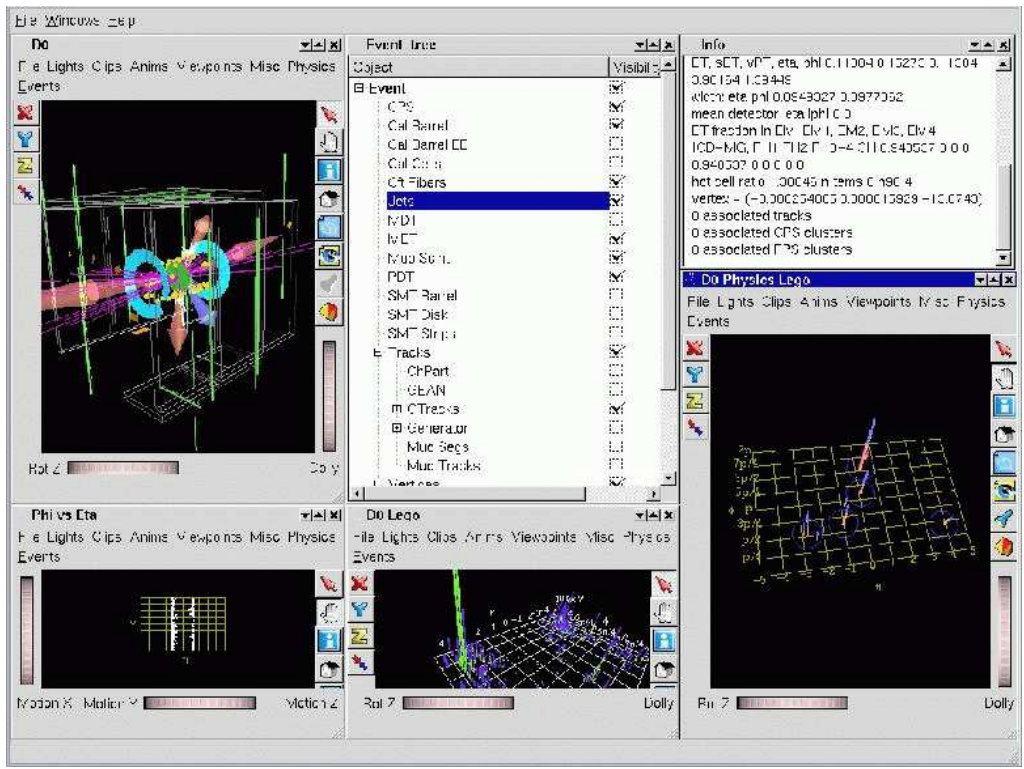}
\caption{D0 event display based on IGUANA.} \label{d0-f2}
\end{figure*}

There are several applications based on IGUANA plug-in
architecture. Among those are OSCAR and ORCA visualisation which are
well known in CMS and actively used for visualising simulated and
reconstructed CMS data. D0Scan (see Figure~\ref{d0-f2}) currently uses
IGUANA as a toolkit. It is used at D0 for event and detector
visualisation.

\subsection{Experiment Specific Plug-ins}

IGUANA is used by other software projects to define
experiment-specific views. Among these projects are the
CMS experiment
projects\footnote{http://cmsdoc.cern.ch/cmsoo/cmsoo.html}: COBRA - core
framework, OSCAR - simulation project, ORCA~\cite{orca} - reconstruction project,
FAMOS - fast simulation project, DDD - detector geometry description
and the D0 experiment project: D0Scan.

The integration with experiment-specific
frameworks is an IGUANA dependent part, usually a sub-system, which is
located within the experiment project. Even though IGUANA core knows
nothing about these sub-systems, the main driver can still load
experiment-specific plug-ins providing that the location of those plug-ins is
listed in the IGUANA\_PLUGINS environment variable as path
defined by a colon-separated list of directories and the
experiment-specific shared libraries are available at run-time.

The projects build plug-ins similar to the described above
(section~\ref{Plgin}). Usually, each project has a dedicated sub-system called
Visualisation for that purpose. The sub-system contains project
specific modules, or plug-ins. The kernel loads the modules
dynamically as required. Currently a user selects what to load at the
interactive setup phase or via a resource file.

The plug-ins are normally demand-loaded shared libraries but
IGUANA also supports statically linked applications, therefore no
shared libraries need to be involved.

\subsection{ORCA Visualisation}

\begin{figure*}[!hbtp]
\centering
\includegraphics[width=60mm]{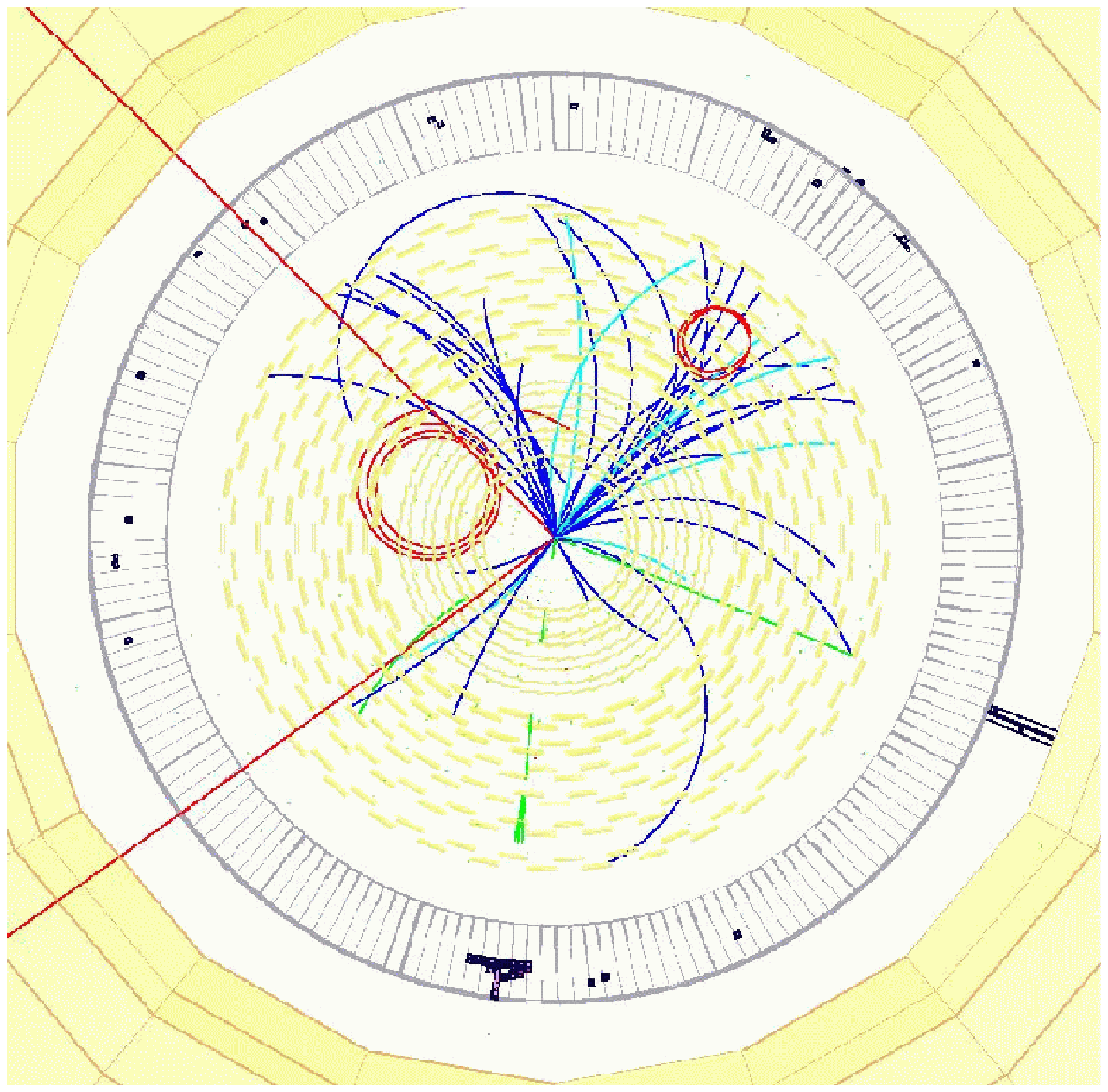}
\includegraphics[width=100mm]{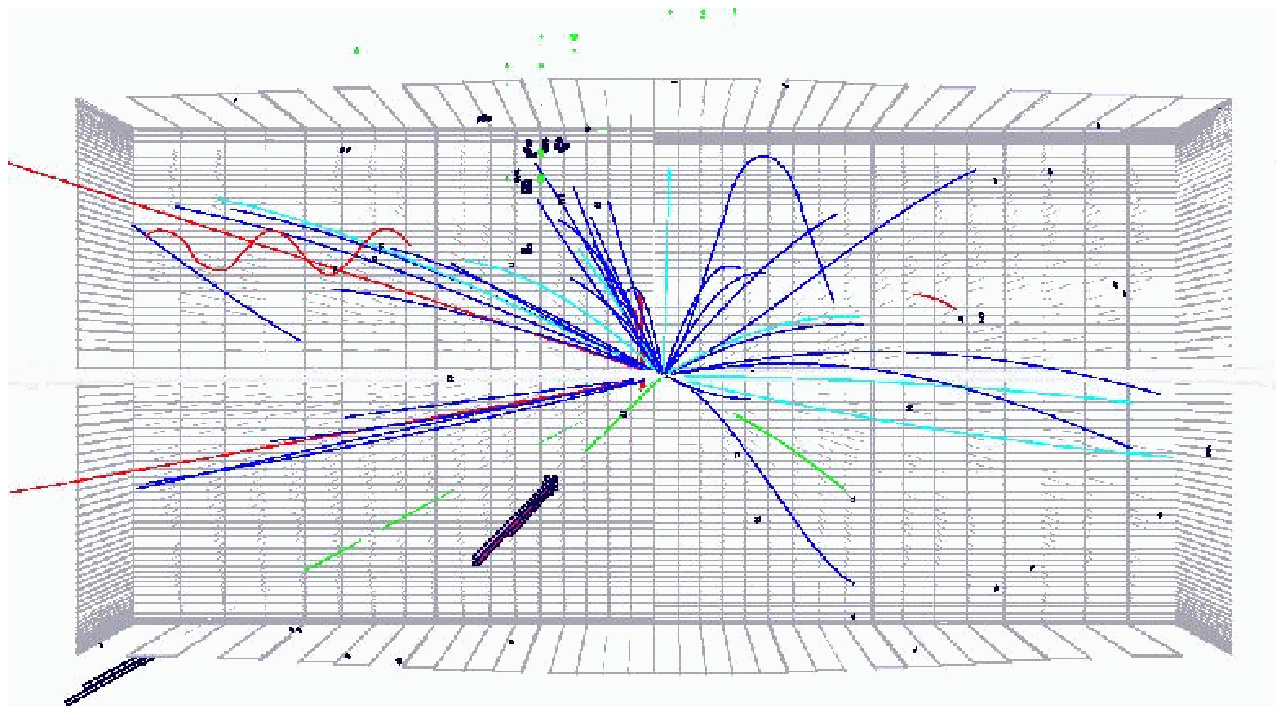}
\caption{ORCA visualisation based on IGUANA: Z view (left) of $H->ee\mu\mu$ event
  and sliced Y view of the same event (right). CMS detector
  description based on Geant3 and reconstructed geometry is shown for
  the tracker silicon detector. Simulated tracks: muons shown with red
  color, electrons - green, pions - blue, all other particles - cyan.} \label{orca-f4}
\end{figure*}

Visualisation for the CMS reconstruction project (ORCA) (see
Figure~\ref{orca-f3} and Figure~\ref{orca-f4})
provides access via COBRA plug-ins to both simulated (cmsim) and
reconstructed event data, and  Geant3 based CMS geometry as well as
reconstructed geometry. Events are requested interactively one by one
from a specified data collection. The IGUANA GUI runs in a separate
thread. Communication between the GUI thread and the event thread is
handled by the COBRA-based event processor plug-in.

The visualisation content is defined at run-time. Plug-ins for
different ORCA sub-systems can be loaded independently. A user can
define the content of the application in an ASCII text resource
file. Various visualisation cuts can also be defined in the resource
file. The on-line configuration editor implemented as an IGUANA
service allows to change them interactively or set new configurables
during the run.

\subsection{D0Scan}

D0Scan is an event viewer for D0 using IGUANA. The D0Scan design is
based upon the IGUANA IgVis module 
which provides a very flexible set of tools for organizing an
OpenInventor scene, creating a GUI to control visibility of items in
the scene, and centralizing  picking callback. Items such as detector
components, hits, tracks, and so forth are implemented as IgQtTwigs
and arranged in a hierarchical structure similar to the OpenInventor
scene graph. This structure is represented in the Twig browser that
controls visibility of graphical objects.

D0Scan makes heavy use of the ability of IgVis to support multiple
windows with different characteristics. User demand has led to the 
development of specialized displays as, for instance, in the
$\eta-\phi$ projection of the distribution of energy in the 
calorimetry.

\subsection{OSCAR Visualisation}


Visualisation for CMS simulation project using GEANT4 - OSCAR
visualisation - is based on the interactive visualisation
sub-framework of IGUANA for GEANT4 described above in
section~\ref{geant4}. CMS detector geometry is described in XML by DDD -
Detector Description Database project. This description is converted
to the GEANT4 one and visualised.

\section{TOOLKIT}


IGUANA makes use of suitable existing software where possible and
integrates the whole in a coherent
fashion. Application developers select from the toolkit only those
parts which are relevant for a particular application.

External software used by IGUANA includes the 3D graphics libraries:
OpenGL and OpenInventor, the Qt toolkit as a GUI kernel, public-domain
Qt and Inventor extensions, various packages for plotting and
interactive analysis, an XML parser, etc.

Being accessible via IGUANA services with dedicated GUI extensions,
both Oprofile and Jprof are used for
debugging. IgNominy~\cite{Ignominy} - an IGUANA
tool for dependency analysis depends on Graphviz for drawing
diagrams. Doxygen is used for auto generated reference
documentation. IGUANA studio printing services integrate gl2ps for
producing high quality vector postscript. The team has improved
culling algorithms of the package which reduce the size of the printed
file dramatically and fed them back to the author.

\subsection{LCG Software and Services}

IGUANA team actively participates in the LCG project. Part of
the software has already migrated to SEAL such as the classlib and
plug-in manager. IGUANA uses services provided by SPI such as Savannah
bug reporting system. POOL is used indirectly via experiment specific
plug-ins.

\section{FUTURE PLANS}

IGUANA will provide wider selection of 2D and 3D representations and
specialized viewers including the time snapshots of visualized data,
algorithm dependent views, and animation.

The project progresses toward coherent physicist desktop
emphasizing the role of wizards for experiment specific environment,
control center, and other services.

Future possibilities are open for discussion in context of the LHC LCG
project, e.g. PI - Physicist Interface project. Among those there are
issues of integration and future relationship with ROOT, JAS, etc. and
potential front-end GUI for GRID applications.

\begin{acknowledgments}
This work is supported by the US National Science Foundation.
\end{acknowledgments}


\begin{thebibliography}{9}   

\bibitem{CMSana}
V. Innocente, ``CMS Data Analysis: Current Status and Future
Strategy'', these proceedings, CHEP03, La Jolla, March, 2003.

\bibitem{Seal}
P. Mato, L. Tuura, et al., ``SEAL: Common core libraries and services for LHC
applications'', these proceedings, CHEP03, La Jolla, March, 2003.

\bibitem{Root}
R. Brun, et al., ``ROOT Status and Future Developments'', these
proceedings, CHEP03, La Jolla, March, 2003. 

\bibitem{Jas}
T. Johnson, et al., ``JAS3 - A general purpose data analysis framework
for HENP and beyond'', these proceedings, CHEP03, La Jolla, March,
2003.

\bibitem{Geant4}
J. Apostolakis, et al., ``An overview of Geant4's recent
developments'', these proceedings, CHEP03, La Jolla, March, 2003.

\bibitem{orca}
V.~Innocente, D.~Stickland, ``Design, Implementation and Deployment of
a Functional Prototype OO Reconstruction Software for CMS. The ORCA
Project.'', Proceedings to CHEP-2000, Padua.\\
http://cmsdoc.cern.ch/ORCA

\bibitem{Ignominy}
L. Tuura, ``Ignominy: Tool for Analysing Software Dependencies and For
Reducing Complexity in Large Software Systems'', Proceedings of the
VIII International Workshop on Advanced Computing and Analysis Techniques
in Physics Research, Moscow, Russia, June, 2002.

\bibitem{arch-01}
G. Alverson, I. Osborne, L. Taylor, L. Tuura ``A Coherent and
Non-Invasive Open Analysis Architecture and Framework with
Applications in CMS'',  Proceedings of CHEP 2001, Beijing, September,
2001. 

\bibitem{iguana-01}
G. Alverson, I. Osborne, L. Taylor, L. Tuura ``The IGUANA Interactive
Graphics Toolkit with Examples from CMS and D0'', Proceedings of CHEP
2001, Beijing, September, 2001.

\bibitem{hippo-01}
P. Kunz, ``The HippoDraw Application and the HippoPlot C++ Toolkit
Upon which it is Built'', Proceedings of CHEP 2001, Beijing,
September, 2001.

\bibitem{scram}
S. Ashby, I. Osborne, J.P. Wellisch, C. Williams, ``Code Organization
and Configuration Management'', Proceedings of CHEP 2001, Beijing,
September, 2001.

\bibitem{Opensci}
G. Barrand, ``Lessons from the Open Scientist Project'', Proceedings
of CHEP 2000, Padova, February, 2000.


\end{thebibliography}

\end{document}